\documentclass[12pt]{article}
 \usepackage{graphicx}
 \usepackage{amsfonts, amsmath,amssymb}
 \numberwithin{equation}{section}
\newcommand\arctg{{\mathrm{arctg\,}}}
 
\begin{document}

 \title{On the theory of Bose√Einstein condensation of quasiparticles: on the
possibility of condensation of ferromagnons at high temperatures}{}%
\author{A.I.~Bugrij${}^{1}$ and V.M.~Loktev${}^{2}$ \\
{\small N.N.~Bogolyubov Institute for Theoretical
Physics}\\ {\small of the National Academy of Sciences of Ukraine}\\
  {\small {\it ul. Metrologicheskaya 14-b, Kiev 03143, Ukraine}}\\
  {\small E-mail: ${}^{1}${abugrij@bitp.kiev.ua},
 ${}^{2}${vloktev@bitp.kiev.ua}}}
 \date{ Fiz.
Nizk. Temp. 33, 51√68 (January 2007)\\(Submitted July 31, 2006)}
 \maketitle
\begin{abstract}
{\small The Bose condensation of magnons in physical systems of finite
size is considered for the case of ferromagnetic thin films. It is shown
that in accordance with present-day experimental capabilities, which
permit one to achieve densities of long-wavelength spin excitations of
$\sim10^{18}$--$10^{19}$ ЯЛ$^{-3}$, in such films, the formation of a
coherent condensate of such quasiparticles begins at temperatures
$T\sim10^{2}$K (including room temperature). It is found that Bose
condensation is accompanied by a scaling phenomenon, according to which
the main thermodynamic variable is not the number of particles $N$ but the
ratio $N/T$. This indicates that the Bose condensation of magnons can be
observed at relatively low magnon densities (and, accordingly, low
pumping). The roles played by the shape of the spectrum of spin
excitations and by the film thickness for observation of the phase
transition to the state with the Bose condensate are analyzed, and the
partial contributions of dierent groups of quasiparticles to the total
spectral distribution of magnons over energies are discussed.}



\end{abstract}

\newpage
\section{Introduction}

 The Bose√Einstein condensation (BEC) of atoms and molecules
\cite{Bose,Ein} has become one of the most remarkable phenomena that
reveal and confirm the quantum nature of a number of macroscopic
processes. The formation of a Bose condensate, i.e., the accumulation of
identical particles with integer spin in one of the quantum states, may be
inherent both to true particles (atoms, molecules) and to quasiparticle
excitations of multiparticle systems. In this sense, quasiparticles ---
excitons and biexcitons and also magnons ≈ are of particular interest,
since, existing only as excited states, they are actually absent (if one
ignores the thermal background) in systems at normal temperature and
pressure. Ordinarily, except under specially chosen conditions, the
equilibrium density of thermal quasiparticles decreases with decreasing
temperature. Therefore, the study of BEC of quasiparticles, or their
``nonthermal'' accumulation in one of the states, requires, first and
foremost, the presence of a macroscopic (essentially nonequilibrium) total
number of quasiparticles. This can be achieved only by using methods of
creating and maintaining a large number of quasiparticles in condensed
systems, at least for a time sufficient for, first, their relatively rapid
thermalization and, second, subsequent BEC. Under such conditions the
latter will occur as a (quasi-)equilibrium phenomenon at a conserved (on
average) number of quasiparticles, which is ensured by some external
source of intense quasiparticle creation.

BEC is a phenomenon associated, as a rule, with very low temperatures: the
critical temperature $T_{\rm BEC}$ at which a Bose condensate is created
depends on the gas density $\mathfrak n$ and  the mass $m$ of its
constituent particles according to the well-known formula
 \cite{Landau}
    \begin{equation}\label{1v}
    T_{\rm BEC}=\frac{{\mathfrak{n}}^{2/3}}{m}
    \frac{2\pi\hbar^{2}}{k_{B}\zeta^{2/3}(3/2)}\,
    \,,\end{equation}
where $\zeta(x)$ is the Riemann $\zeta$-function.  For example, in
experiments on the observation of BEC \cite{Anderson,Davis} because of the
large mass of the atoms of alkali elements and low density (approximately
$10^{3}$ particles in a volume of $\sim10^{-6}\ {\text{ЯЛ}}^{3}$) the
system must be ``cooled'' down to $10^{-9}\ldots10^{-8}$K. Or, for
example, in the case of the ``most ideal`' of the real gases, helium, the
temperature of the transition to the Bose-condensed state has a value
$T_{\rm BEC}\approx2\times10^{-2}$й at a density $\approx10^{19} {\text
{ЯЛ}}^{-3}$, which corresponds to a gaseous state. It is therefore obvious
that observation of BEC at high (up to room) temperatures is possible only
in systems consisting of light (weakly interacting) bosons. As has been
mentioned repeatedly, quasiparticle excitations, the effective mass of
which is rather small --- in particular, comparable to the electron mass
$m_{e}$ --- are well suited to the role of such bosons.

In the discussion of the BEC of quasiparticles, attention has apparently
been focused primarily on excitons and biexcitons. Starting with the
pioneering work of Moskalenko \cite{Mosk} and Keldysh \cite{Keldysh}, many
papers have been devoted to the study of the phase diagram of
semiconductors at high densities of nonequilibrium electrons and holes and
also to attempts to observe different types of their joint condensed
states. Despite the short lifetime of excitonic states and the intense
processes of electron√hole annihilation, there have been several pieces of
convincing experimental evidence in favor of the observation of exciton
(and biexciton) BEC \cite{McDonald}.

However, there are other promising systems for studying the BEC of
quasiparticles. In particular, among the interesting objects suitable for
this (see Ref. \cite{Kagan}) are magnets, including ferromagnetic
insulators, where high densities of magnons (up to $10^{18}$--$10^{19}
{\text {ЯЛ}}^{-3}$) can be created by pulsed microwave pumping. We recall
that magnons, which are elementary excitations above the magnetization
field, to good approximation obey Bose statistics\footnote[1]{
Interestingly, it was stated in Ref. \cite{McDonald} that all known cases
of BEC involve the concept of composite bosons formed by an even number of
fermions. This pertains to both particles ≈ helium and alkalimetal
atoms≈and to quasiparticles≈Cooper pairs, excitons, and biexcitons. This
is undoubtedly true if one is talking about particles or about
Wannier√Mott excitons and biexcitons. It is harder to agree with such a
statement for small-radius excitons, or Frenkel excitons. And it clearly
does not apply to such quasiparticles as magnons, especially in insulators
describable by the Heisenberg model.} Their spectrum is formed on account
of the presence of several interactions, chief among which are the
exchange and magnetic dipole√dipole interactions \cite{Gur}. The first of
these determines an isotropic spectrum of spins waves with wavelengths
$k^{-1}$, less than the characteristic size $L$ of the system, so that
$kL\gg1$. The second, on the contrary, operates at wavelengths for which
$kL\leq1$ and, furthermore, leads to dependence of the magnon spectrum on
the direction of $\mathbf{k}$. Analysis of the conditions for BEC of
specifically these elementary excitations in comparatively thin slabs
(microfilms) is the subject of the present study, which is directed toward
the task of examining the conditions for observation and the features of
BEC in ferromagnetic  insulating films with different shapes of the
spin-wave spectrum. We consider magnons to be very promising for research
on the BEC of quasiparticles for at least several reasons. First, they
have a relatively long lifetime and, as we have said, can have densities
reaching $\sim10^{19} {\text {ЯЛ}}^{-3}$; second, there are technologies
in place for growing very thin (less than $\lesssim10\, \mu m$ thick)
films with rather perfect structure and for making precision measurements
of the magnon spectrum in them by optical methods; third, the temperature
and magnetic field can be easily varied over wide limits, permitting
detailed comparison of theory and experiment.

We note that in recent years a rather large number of papers have appeared
(e.g., Refs. \cite{Nik}--\cite{Crisan}) in which the idea of BEC of
magnons has been used for describing phase transitions in (predominantly)
antiferromagnets from their nonmagnetic (singlet) state to a magnetically
ordered state under the influence of an external magnetic field. The point
is that the induced appearance of magnetization in finite fields can
actually be formally described and interpreted in the language of
condensation of magnetic excitations. However, BEC as such does not occur,
since there one is talking about only a rearrangement of the ground state
of the system and, consequently, of virtual and not real magnons (see,
e.g., Ref. \cite{De1}). In contrast, the focus of our attention here is on
real --- excited --- states that appear as a consequence of pumping
electromagnetic energy into the system, this energy going to the creation
of quasiparticles above the ground state, as was masterfully done in a
recent experimental study with the use of microfilms of yttrium iron
garnet (YIG) \cite{Demo}.

\section{Model and general relations}
Suppose that we have a ferromagnetic crystal in the form a parallelepiped
(see Fig.~1) of volume $V=L_{x}L_{y}L_{z}$.
\begin{figure}[h] \begin{center}
\includegraphics[height=55mm,keepaspectratio=true]
{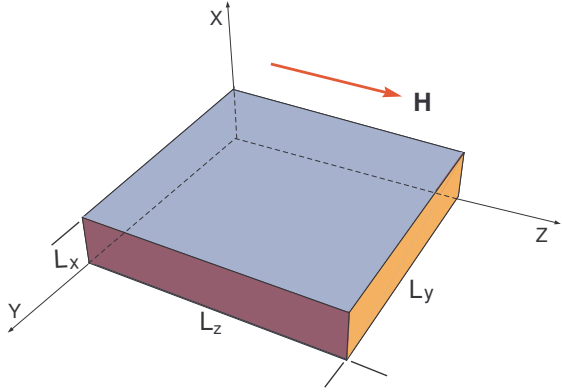} \caption{\small  Shape of the ferromagnetic crystal. }
 \end{center} \end{figure}
 The number of sites $N_{j}$ along
the corresponding axes ($j$ takes the values $x,\ y,\ z$) is determined by
the lattice parameters $a_{j}$: $N_{j}=L_{j}/a_{j}+1$. As we know, the
magnetic ground state of such a crystal presupposes an identical direction
of all the spins (e.g., along the quantization axis $z$), while the lowest
excited state corresponds to one ``flipped'' spin \cite{Dav,Bar}. From a
linear combination of such states one can construct an eigenstate in the
form a spin wave with amplitude $b_{\mathbf{q}}(\mathbf{r})$. The form of
these amplitudes depends on the boundary conditions. Usually for
simplification one uses {\it cyclic} boundary conditions, when the
amplitudes are identical at opposite faces of the crystal. Then the
amplitudes have the form of a plane wave:
\begin{equation}\label{1d2}
b_{\mathbf{q}}(\mathbf{r})=\exp(i\mathbf{q}\mathbf{r})\,\prod_{j}N_{j}^{-1/2}\,,\end{equation}
where in the given case the vector  $\mathbf{r}$ enumerates lattice sites
--- $r_{j}=1,\,2,\,\ldots,\,N_{j}$, and the
dimensionless quasimomentum $\mathbf{q}$ is defined in the Brillouin zone
$(-\pi<q_{j}\leq\pi)$ and takes on a discrete spectrum of values with a
step $\Delta q_{j}^{per}=2\pi/N_{j}$.

If, on the contrary, the spins on the faces are ``free'', the solution of
the boundary condition problem with the corresponding (free) boundary
conditions leads to amplitudes that differ from (\ref{1d2}):
\begin{equation}\label{2d2}
b_{\mathbf{q}}(\mathbf{r})=\prod_{j}\sin(q_{j}r_{j})[2/(N_{j}+1)]^{1/2}
\,.\end{equation}

The Brillouin zone here is defined somewhat differently: $0<q_{j}<\pi$,
and the discreteness step is $\Delta q_{j}^{free}=\pi/(N_{j}+1)$. Let us
point out the main differences between the quasimomentum spectra in
(\ref{1d2}) and (\ref{2d2}). First, for periodic boundary conditions there
exists a zero mode  $\mathbf{q}=0$, Юwhile for the free boundary
conditions there is not: the minimum value that can be taken on by each of
the quasimomentum components is $q_{j}^{min}=\Delta q_{j}^{free}$. Second,
the quasimomentum spectrum for free boundary conditions is twice as dense
as for periodic boundary conditions, since for $N_{j}\gg1$ the ratio
$\Delta q_{j}^{per}/\Delta q_{j}^{free}=2N_{j}/(N_{j}+1)\approx2$.

It is considered to be almost obvious that the concrete form of the
boundary conditions does not influence the behavior of physical
quantities. Actually this is true if one is talking about very ``large''
systems. However, as will be shown below, for ``small'' systems the
difference mentioned above --- the absence of a zero mode in the spectrum
for the case of free boundary conditions --- leads to quite perceptible
contributions (Weyl corrections) to certain observable characteristics
which are absent in the case of idealized periodic boundary conditions.
Since we intend to investigate the question of BEC in microfilms $\sim1
\mu m$ thick, the use of free boundary conditions is justified as more
realistic.

In spite of the differences indicated, the expression for the magnon
energy due to the isotropic exchange interaction is the same for both
kinds of boundary conditions and can be written in dimensionless form as
  \begin{equation}\label{1d}
 \epsilon(\mathbf{q})=2\sum_{j}\sin^{2}\frac{q_{j}}{2}\,. \end{equation}
Far from the Brillouin zone boundary or in long-wavelength region
$(q_{j}\ll\pi)$, it follows from (\ref{1d})) that
  \begin{equation}\label{2d}
 \epsilon(\mathbf{q})=\frac{\mathbf{q}^{2}}{2}\,.\end{equation}

The dimensional quantities are easily restored through the dimensions of
the crystal $L_{j}$, the lattice parameters $a_{j}$, and the effective
mass $m_{m}$, which for magnons is inversely proportional to the exchange
integral. Then the components of the quasimomentum for the case of free
boundary conditions
 $$p_{j}=\hbar\,\frac{q_{j}}{a_{j}}\approx\hbar\pi\frac{k_{j}}{L_{j}},\quad
 k_{j}=1,2,\ldots,N_{j},$$ while the dispersion relation
 \begin{equation}\label{4d}
 \varepsilon(\mathbf{p})=\frac{\mathbf{p}^{2}}{2m_{m}}=
 \sum_{j}\varepsilon_{j}\,k_{j}^{2}
 \equiv\varepsilon_{\mathbf{k}},\quad
 \varepsilon_{j}=\frac{\hbar^{2}\pi^{2}}{2m_{m}L_{j}^{2}}\end{equation}
corresponds to the simplest one for magnons in a ferromagnet. Also taking
into account the fact that the long-wavelength magnons interact weakly,
$\sim(\mathbf{p}_{1}\mathbf{p}_{2})^{2}$, with each other and (with
practically the same amplitude) with phonons\cite{Bar}, we arrive at the
problem of BEC of an ideal\footnote[2]{The ideality is determined by the
density of Bose excitations, which even for a quasiparticle number
$\sim10^{20}{\text{ЯЛ}}^{-3}$ turns out to be extremely small:
$\sim10^{-3}$ per atom.} Bose gas, for which the partition function
$Z_{m}$ in the grand canonical ensemble at temperature $T$ and chemical
potential $\mu_{m}$ has the form
 \begin{equation}\label{5d} \ln Z_{m}
=-\sum_{\mathbf{k}}\ln\bigl\{1-\exp[-(\varepsilon_{\mathbf{k}}-\mu_{m})/T]
\bigr\}\,,\end{equation} where $\varepsilon_{\mathbf{k}}$ --- the energy
of the state with quantum numbers $\mathbf{k}=(k_{x},k_{y},k_{z})$, is
defined in (\ref{4d}). In (\ref{5d}) and below we employ the system of
units $k_{B}=\hbar=1$, restoring the dependence on the fundamental
constants as necessary; in addition, the index m will be dropped with the
understanding that the particles under discussion are magnons. Then for
the mean number of particles in quantum state ${\mathbf{k}}$ (occupation
number) we have
\begin{equation}\label{6d}
n_{\mathbf{k}}=\{\exp[(\varepsilon_{\mathbf{k}}-\mu
)/T]-1\}^{-1}\,,\end{equation} and for the total (mean) number of
particles in the system \begin{equation}\label{7d} N =T\frac{\partial\ln Z
}{\partial\mu }=\sum_{\mathbf{k}}n_{\mathbf{k}}.\end{equation}

It follows from the domain of definition of the thermodynamic quantities
(\ref{5d})--(\ref{7d}) that the range of variation of the chemical
potential for Bose systems is bounded by the minimum value of the energy
in the dispersion relation under study. In the case of dispersion relation
(\ref{4d}) the energy reaches a minimum value $\varepsilon_{0}$ in the
quantum state with $\mathbf{k}=(1,1,1)$:   $$
\varepsilon_{0}=\frac{\pi^{2}}{2m }\sum_{j}L_{j}^{-2}\,, \quad \mu
<\varepsilon_{0}\,.$$ We note, by the way, that the widespread assertion
that the chemical potential is equal to zero as a consequence of the free
creation and annihilation of (quasi)particles is not completely correct.
This error apparently stems from the fact that the free energy $F(N,T)$ in
the canonical ensemble reaches a minimum as a function of the number of
particles (at fixed volume and temperature) at $\mu=0$. This is
essentially a trivial consequence of one of the definitions of the
chemical potential as a thermodynamic function, $\mu=\partial F/\partial
N$.  But why is the condition of minimum free energy equivalent to the
condition of free creation and annihilation of particles? Logically it
would seem to flow from the requirement of maximum entropy as a function
of $N$. Meanwhile, it is easy to check by elementary calculations (in the
case of an ideal gas, at least) that the entropy is a monotonically
increasing function of the number of particles and reaches its maximum
value at $N\to\infty$. Analogously, in the grand canonical ensemble the
entropy is a monotonically increasing function of the chemical potential
and reaches its maximum at the maximum admissible value of the latter,
i.e., when $\mu\to\varepsilon_{0}$. Here both the mean number of particles
$N$ and the entropy itself go to infinite in {\it a finite volume}. The
chemical potential $\mu$ as an independent thermodynamic variable,
although a convenient parameter for theorists, is absolutely a formal
quantity from the standpoint of experimentalists, since there is no
prescription for its direct measurement. In experiment the chemical
potential can only be inferred indirectly, e.g., from measurement of the
mean number of particles or other observables. Then the value of $\mu$ can
be recovered on the basis of some theoretical prescriptions: as a rule,
from the formulas of statistical mechanics of an ideal gas. In the case of
an ideal gas, however, one can eliminate the chemical potential completely
from the thermodynamic formulas by replacing this independent
thermodynamic variable by some other quantity having a more direct
physical content. If one is talking about BEC, a completely suitable
candidate for this role \cite{Fej} is the number of particles at the
lowest level, $n_{0}$. In fact, it follows directly from Eq. (\ref{6d})
that
\begin{equation}\label{9d} \mbox{e}^{(\varepsilon_{0}-\mu
)/T}=1+1/n_{0}\,.\end{equation} Then, substituting (\ref{9d}) into
(\ref{5d})--(\ref{7d}), we obtain
\begin{eqnarray}\label{10d} \ln Z
&=&\ln(n_{0}+1)-{\sum_{\mathbf{k}}}'\ln\biggl[1-
\mbox{e}^{-(\varepsilon_{\mathbf{k}}-\varepsilon_{0})/T}\frac{n_{0}}{n_{0}+1}
\biggr]\,,\\\label{11d} n_{\mathbf{k}}&=&\bigl[(1+n_{0}^{-1})\,
\mbox{e}^{(\varepsilon_{\mathbf{k}}-\varepsilon_{0})/T}-1\bigr]^{-1}\,,\\
\label{11d1} N
&=&n_{0}+{\sum_{\mathbf{k}}}'n_{\mathbf{k}}\,,\end{eqnarray} where the
prime on the summation sign means that the term corresponding to the
lowest magnon state with quantum numbers $\mathbf{k}=(1,1,1)$ is excluded.

Such a parametrization is convenient for two reasons. First, it gives a
formal definition of the condensate in a finite system as simply a set of
particles at the lowest-energy quantum level: the number of these
particles is $N_{\rm BEC}=n_{0}$. Furthermore, it permits a correct
transition to the thermodynamic limit in expressions obtained for systems
of finite size. This question will be discussed in a little more detail
when we specialize to microfilms.

Of course, in the sums over quantum states in (\ref{5d}), (\ref{7d}) and
(\ref{10d}), (\ref{11d1}) the dependence on the size of the system enters
implicitly ≈ through the spectrum of the Schr\"odinger operator, which
depends on the boundary conditions. To isolate this dependence we resort
to the standard technique of changing from summation over quantum states
to integration over the phase volume:
 \begin{equation}\label{12d} \sum_{\mathbf{k}}f(p_{\,\mathbf{k}})\rightarrow\int
d\Phi f(p)\,,\end{equation} where the element of phase volume is usually
taken as
 \begin{equation}\label{13d}
 d\Phi=\frac{V}{(2\pi\hbar)^{3}}\,d^{3}p=
 \frac{V}{2\pi^{2}\hbar^{3}}\,p^{2}\,dp\equiv d\Phi_{V}\,.\end{equation}
 It should not be forgotten, however, that expression (\ref{13d}) is only the first term of an
asymptotic expansion for $p\to\infty$. According to the famous Weyl
problem \cite{Weyl} of the number of eigenvalues of an operator which do
not exceed a specified value, the coefficients of the corresponding
asymptotic series are expressed in terms of geometric invariants. In the
particular case of the Schr\"odinger operator, taking the next term into
account leads to the following expression for the phase volume:
\begin{eqnarray}\nonumber d\Phi&=&d\Phi_{V}-d\Phi_{S_{V}}\,,\\\label{14d}
 d\Phi_{S_{V}}&=&\frac{S_{V}}
 {8\pi\hbar^{2}}\,p\ dp\,,\end{eqnarray} where $S_{V}$ is the surface area of the sample of volume $V$.
 It was mentioned above that the
contribution $d\Phi_{S_{V}}$ is due precisely to the absence of the zero
mode in bounded systems. It is not difficult to see that integration over
the phase volume elements (\ref{13d}) and (\ref{14d}) generates the
quasiclassical expansion for the corresponding thermodynamic variables.
Transformation (\ref{12d}) also presupposes that the function $f(p)$ in
the integrand is rather smooth. Otherwise some terms (say, the first) can
differ sharply in value from the others. Then one needs to separate them
off and apply transformation (\ref{12d}) to the remaining sum. A simple
example illustrating the difference between the sum and integral in such a
situation is given in Appendix A. The partition function in the BEC
regime, i.e.,    for $\mu \to\varepsilon_{0}$, is the case that was
indicated in \cite{Landau}, for example. It is seen that the change of
independent thermodynamic variable (\ref{9d}) from the chemical potential
to $n_{0}$ solves in one stroke the problem of separating out the singular
contributions [cf. (\ref{5d}), (\ref{7d}) and (\ref{10d}), (\ref{11d1})].
The corrections to the thermodynamic variables due to $d\Phi_{S_{V}}$ we
will call {\it surface} contributions, and the contributions due to
$d\Phi_{V}$,
 {\it volume} contributions. We note that their ratio is not universal, in the
sense that it can be larger or smaller for different thermodynamic
functions under equal conditions. In other words, as the volume of the
system increases, the thermodynamic limit sets in sooner for some physical
quantities and later for others. The study of rather thin films is clearly
a step in the direction of the physics of mesoscopic systems, or
``nanophysics'', as it has come to be called. As a quantitative estimate
of the boundary between ``macroscopic'' and ``mesoscopic'' one can use a
comparison of the volume and surface contributions to a given physical
quantity. When they are comparable, the formulas of ordinary macroscopic
thermodynamics cease to work. In particular, the energy, free energy,
entropy, and even the mean number of particles lose the property of
extensivity and, for example, such an interesting quantity as the pressure
loses the property of isotropy.

Turning to concrete calculations, we note that averaging over the phase
volume (\ref{14d}) leads to integrals of the form
\begin{equation}\label{116d}\int\limits_{0}^{\infty}\frac{dy\
y^{\alpha-1}}{\mbox{e}^{y+x}-1}=\Gamma(\alpha)P_{\alpha}(\mbox{e}^{-x})\,,\end{equation}
where  $\Gamma(\alpha)$ is the gamma function, and $P_{\alpha}(z)$ is a
polylogarithm, a special function with rather simple properties. In the
region ${\rm Re}\, x>0$ it has the series expansion
\begin{equation}\label{117d}P_{\alpha}(\mbox{e}^{-x})=
\sum_{l=1}^{\infty}l^{-\alpha}\mbox{e}^{-lx}\,,\end{equation} which
implies that
\begin{equation}\label{118d}\frac{d}{dx}P_{\alpha}(\mbox{e}^{-x})=-P_{\alpha}(\mbox{e}^{-x})\,.
\end{equation} At $x=0$ there is a branch point $P_{\alpha}(\mbox{e}^{-x})\sim x^{\alpha-1}$
 for noninteger $\alpha$, and $P_{\alpha}(\mbox{e}^{-x})\sim x^{\alpha-1}\ln x$
for integer $\alpha$; this branch point is explicitly separated out. For
example, the function $P_{5/2}(\mbox{e}^{-x})$ has the representation
\begin{equation}\label{119d}P_{5/2}(\mbox{e}^{-x})=
\frac{4}{3}\sqrt{\pi}\,x^{3/2}+\sum_{l=0}^{\infty}
\frac{\zeta(5/2-l)}{l!}(-x)^{l}\,,\end{equation} where the radius of
convergence of the series on the right-hand side of where the radius of
convergence of the series on the right-hand side of (\ref{119d}) is
$|x|<2\pi$; as we have said, $\zeta(l)$ is the Riemann $\zeta$ function.

Let us calculate, for example, the mean number of particles (\ref{12d})
 \begin{eqnarray} \nonumber N
&=& n_{0}+N_{\rm ex},\\\nonumber N_{\rm
ex}&=&{\sum_{\mathbf{k}}}'n_{\mathbf{k}}=N_{V}-N_{S_{V}},\\\label{120d}
N_{V}&=&\frac{V}{\lambda_{T}^{3}}P_{3/2}\biggl(\frac{n_{0}}{n_{0}+1}
\biggr)\,,\\\nonumber N_{S_{V}}&=&\frac{S_{V}}{4\lambda_{T}^{2}}P_{1}
 \biggl(\frac{n_{0}}{n_{0}+1}\biggr)=\frac{S_{V}}{4\lambda_{T}^{2}}
 \ln(n_{0}+1)\,,
\end{eqnarray} where   \begin{equation}\label{121d} \lambda_{T}=
\hbar\,\sqrt{\frac{2\pi}{m k_{B}T}}\end{equation} is the so-called thermal
de Broglie wavelength. It follows from (\ref{120d}) that at a temperature
of 1~K, quasiparticle mass $m_{e}$, and number of particles in the
condensate $n_{0}\sim10^{16}$, the contributions $N_{V}$ and $N_{S_{V}}$
become equal when the film thickness decreases to $L_{x}\sim1\,\mu m$,
which, in accordance with what we have said above, is a direct indication
of mesoscopicity of the system.

The problem of BEC of an ideal gas is one of the few that have an exact
solution for a large (but finite) number of particles. The solution
permits one to trace the formation of nonanalyticity of some physical
quantity or other as a function of temperature at the transition to the
thermodynamic limit and to find a quantitative estimate of this transition
to the limit. Therefore, to complete the general picture, let us discuss
the question of BEC as a phase transformation phenomenon. The order
parameter here is the condensate density ${\mathfrak n}_{\rm
BEC}=n_{0}/V$: ${\mathfrak n}_{\rm BEC}=0$ for $T>T_{\rm BEC}$, and
${\mathfrak n}_{\rm BEC}\sim1-T/T_{\rm BEC}$ for $T\lesssim T_{\rm BEC}$.
The order of the phase transition is determined in a suitable
classification according to whether a jump of the derivative of the heat
capacity occurs upon transition of the temperature through $T_{\rm BEC}$.

Expressions for the number of particles and energy are written without the
surface terms as
\begin{eqnarray} \label{110d} N &=&n_{0}+{\sum_{\mathbf{k}}}'n_{\mathbf
{k}}=n_{0}+\frac{V}{\lambda_{T}^{3}}\,P_{3/2}\biggl(\frac{n_{0}}{n_{0}+1}
\biggr)\,,\\
\label{111d} E &=&n_{0}\varepsilon_{0}+{\sum_{\mathbf{k}}}'n_{\mathbf
{k}}\varepsilon_{\mathbf
{k}}=n_{0}\varepsilon_{0}+\frac{3}{2}\frac{TV}{\lambda_{T}^{3}}\,P_{5/2}
\biggl(\frac{n_{0}}{n_{0}+1} \biggr)\,.\end{eqnarray} It follows from the
definition (\ref{1v}) of the temperature $T_{\rm BEC}$ that
\begin{equation} \label{112d} \frac{V}{\lambda_{T}^{3}}=\frac{N
}{\zeta(3/2)}\,.\end{equation} Introducing a normalized temperature
$t=T/T_{\rm BEC}$ and taking (\ref{112d}) into account, we rewrite
(\ref{110d}) and (\ref{111d}) in the form
\begin{eqnarray} \label{113d} N &=&n_{0}+\frac{N
t^{3/2}}{\zeta(3/2)}\,P_{3/2}\biggl(\frac{n_{0}}{n_{0}+1}
\biggr)\,,\\
\label{114d} \frac{E }{T_{\rm BEC}} &=&\frac{n_{0}\varepsilon_{0}}{T_{\rm
BEC}}+\frac{3}{2}\frac{N t^{5/2}}{\zeta(3/2)}\,P_{5/2}
\biggl(\frac{n_{0}}{n_{0}+1} \biggr)\,.\end{eqnarray}  It is seen from
(\ref{113d}) that at temperatures in the vicinity of $T_{\rm BEC}$
$(t\approx1)$ and under the condition $N \gg1$, the number of particles in
the BEC is also large: $n_{0}\sim N ^{2/3}$, both above and below $T_{\rm
BEC}$; in particular, for $t=1$
\begin{equation}\label{222d}
n^{3/2}_{0}=\frac{2\sqrt{\pi}\,N}{\zeta(3/2)}\, [1-O(N
^{3/2})].\end{equation} For the specific heat $c$ and its derivative at
constant volume $V$, we obtain from (\ref{113d}) and (\ref{114d})
\begin{equation}\label{20d} c =\frac{1}{N T_{BEC}}\biggl(\frac{\partial
E}{\partial t}+\frac{1}{u}\frac{\partial E}{\partial n_{0}}\biggr)\,,\quad
\frac{dc }{dt}=\frac{\partial c }{\partial t}+\frac{1}{u}\frac{\partial c
}{\partial n_{0}}\,,\end{equation} where \begin{equation}\label{21d}
u\equiv\frac{dt}{dn_{0}}=\frac{1}{N -n_{0}}+\frac{1}{n_{0}(n_{0}+1)}
\,\frac{P_{1/2}[n_{0}/(n_{0}+1)]}{P_{3/2}[n_{0}/(n_{0}+1)]}\,.\end{equation}
For
 $N\gg n_{0}\gg1$  \begin{equation}\label{23d} u=\frac{1}{N
}+\frac{\sqrt{\pi}}{\zeta(3/2)}n_{0}^{-3/2} +O(N ^{-4/3})\,.\end{equation}
This last expression shows that the scaling variables\footnote[3]{By
scaling behavior we mean that some function (e.g., of two variables)
$f(x,s)$, starting at certain scales $s>s_{0}$ degenerates into a function
of only one, ``scaling'' variable $y$:
$f(x_{1},s_{1})=f(x_{2},s_{2})=\tilde{f}(y)$, where $y=y(x,s)$.  In the
problem under consideration the scale is the size of a system  $s\sim
L\sim N^{1/3}$.}  of the problem are
\begin{equation}\label{24d} \tau=N (t-1),\quad
y=\frac{\zeta(3/2)}{\sqrt{\pi}}\,\frac{n_{0}^{3/2}}{N }\,.\end{equation}
The asymptotic expansions for (\ref{113d}), (\ref{20d}), and (\ref{21d})
take the form
\begin{eqnarray} \nonumber \tau&=&\frac{2}{3}\ \frac{1-y}{y^{1/3}}+O(N
^{-1/3}),\\\nonumber c
&=&c_{max}\biggl[1-\frac{\alpha}{(yN)^{1/3}}(y+\frac{2\gamma}{1+y}-1)
\biggr]+O(N ^{-2/3}),\\\label{25d} c' &\equiv&\frac{dc
}{dt}=\frac{3}{2}c_{max}\biggl[1-\gamma\frac{1+4y}{(1+y)^{3}} \biggr]+O(N
^{-1/3}),\end{eqnarray} where
$$c_{max}\equiv\frac{15}{4}\frac{\zeta(5/2)}{\zeta(3/2)}\,,\quad
\alpha=\frac{\sqrt{\pi}}{\zeta(3/2)}\,,\quad
\gamma=\frac{3}{10\pi}\,\frac{\zeta^{3}(3/2)}{\zeta(5/2)}\,.$$ It is seen
from (\ref{25d}) that the small parameter in these expressions is $N
^{-1/3}$. Consequently, the average of physical quantities tends toward
the thermodynamic limit rather slowly --- according to an $O(V^{-1/3})$
law rather than $O(V^{-1})$ as is usually assumed. The scaling regime sets
in even more slowly, as is illustrated in Fig.~2.
\begin{figure}[h] \begin{center}
 \includegraphics[height=60mm,keepaspectratio=true]
 {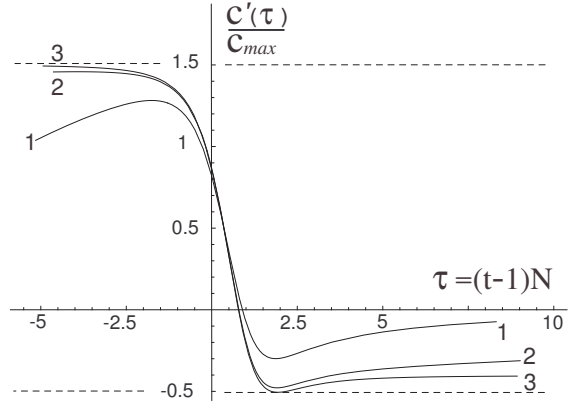}
 \caption{\small \ Derivative of the heat capacity as a function of the
 (scaled) temperature for different
numbers of particles in the system: $N=10^{3}$ (1), $N=10^{6}$ (2),
$N\to\infty$ (3).}
 \end{center} \end{figure}

At arbitrarily low but fixed deviations of the temperature from $T_{\rm
BEC}$ ($t=1+\delta t$) the variable $y$ (\ref{24d}) has substantially
different dependence on $N$ for different signs of the deviation $\delta
t$: $$y=\frac{1}{N }\,\frac{1}{\delta t^{3}}\ll1,\quad \delta t>0,$$
$$y=-N ^{1/3}\delta t\gg1,\quad \delta t<0.$$ Accordingly, from (\ref{25d})
 we obtain the following
expression for the jump of the derivative of the specific heat $\Delta c'
=c' (1-\delta t)-c' (1+\delta t)$ at $N \to\infty$, $\delta t\to0$:
$$\Delta c' =\frac{3}{2}\,c_{max}\,\gamma=\frac{9}{16\pi}
\zeta^{2}(3/2),$$ which agrees with the known expression given in
\cite{Landau}. Returning to the ordinary, unnormalized temperature, we can
see that the function $c' $ for $N \to\infty$ goes over to a $\theta$
function with a ``tooth'' formed by the minimum on the curve (Fig.~2).

Based on the given calculations, we stress that for BEC as a phase
transformation phenomenon it is also necessary to have not too low
temperatures at a fixed density of the Bose gas, so as to have a large
number of (quasi)particles in the system\footnote[4]{In this connection we
note that an atomic Bose condensate with a total number of particles
$\sim10^{3}$ (Refs. \cite{Anderson} and \cite{Davis}), while qualitatively
corresponding to the BEC phenomenon, nevertheless has little in common
with the original and widely accepted conception of this phenomenon.}
Therefore, for experimental study of BEC it is preferable to increase the
density of excitations at a specified value of $T$, as is done in Ref.
\cite{Demo}.

\section{Bose condensation in thin films with anisotropic spin-wave
spectrum} A study of magnons in ferromagnetic microfilms with dimensions
\cite{Demo}  \begin{equation}\label{1a1} L_{x}\simeq10 \mu m,\quad L_{y}=
L_{z}\simeq1 {\text {ЯЛ}}\end{equation} in a magnetic field directed along
the surface of the sample, $\mathbf{H}||z$ , was reported in Ref. 18 (see
Fig.~1). The spectrum of long-wavelength spin excitations in real
ferromagnets is rather complex. The corresponding dispersion relation
$\varepsilon(\mathbf{p})$ differs markedly from (\ref{4d}), which
describes free massive particles. The additional contributions to
$\varepsilon(\mathbf{p})$ are generated by an external magnetic field and
the magnetic dipole component of the interaction between spins. In the
next Section the influence of these contributions on the BEC process are
analyzed in more detail. In the simplest approximation one may keep only
the fact that a magnetic-field gap appears in the dispersion relation, so
that
\begin{equation}\label{1a2}
\varepsilon(\mathbf{p})=\frac{\mathbf{p}^{2}}{2m}+\varepsilon_{\mathbf{H}}
\,,\end{equation} where
$\varepsilon_{\mathbf{H}}=2\pi\hbar\nu_{\mathbf{H}}$ is the gap energy,
its frequency practically coinciding with the ferromagnetic resonance
(FMR) frequency. The presence of a gap in the dispersion relation does not
affect the thermodynamics of the system, since this is only a shift of the
energy (frequency) scale. As is easily seen from the definition (\ref{9d})
of $n_{0}$, the substitution
$\varepsilon\to\varepsilon+\varepsilon_{\mathbf{H}}$ and
$\mu\to\mu+\varepsilon_{\mathbf{H}}$ leaves $n_{0}$, the number of
particles on the lowest level, which essentially corresponds to the FMR,
practically unchanged. And because it is $n_{0}$, not the chemical
potential, that we use as the independent variable, all of the
thermodynamic expressions of the previous Section remain in force. The
quantity $\varepsilon_{\mathbf{H}}$ is conveniently used as a scale for
parameters with dimensions of energy, if only because the diagnostics of
the spectral density of of magnons in YIG in \cite{Demo} was carried out
at frequencies $\nu_{\mathbf{H}}\simeq2\,{\rm GHz}$, in the vicinity of
the FMR. Then, e.g., the temperature is given as
$T=t/\varepsilon_{\mathbf{H}}$, which implies that, on the scale of
quantities corresponding to those typical of \cite{Demo}, a temperature of
1~й corresponds to a value $t\simeq10$. In the dimensionless quantities
dispersion relation (\ref{1a2}) takes the form
 \begin{equation}\label{1}
\varepsilon(\mathbf{p})\equiv\varepsilon_{\mathbf{k}}=
\varepsilon_{\mathbf{H}}(1+\omega_{\mathbf{k}})\,,\quad
\omega_{\mathbf{k}}=\sum_{j}\omega_{j}k_{j}^{2}\,,\end{equation} where
$\omega_{j}=\varepsilon_{j}/\varepsilon_{\mathbf{H}}$, the
$\varepsilon_{j}$
  being defined in (\ref{4d}). In the case of film dimensions (\ref{1a1})
and magnon masses $m_{m}\simeq5m_{e}$, corresponding to YIG \cite{Gur},
the numerical values of the parameters $\omega_{j}$ are
\begin{equation}\label{2} \omega_{x}\simeq10^{-4},\quad
\omega_{y}=\omega_{z}\simeq10^{-10}.\end{equation}

In the experiments of \cite{Demo}, direct measurements were made of the
magnon spectral density  \begin{equation}\label{3}
n(\omega)=\sum_{\mathbf{k}}n_{\mathbf{k}}\delta(\omega-\omega_{\mathbf{k}})
=\frac{g (\omega)}{\mbox{e}^{(\omega-\omega_{0})/t}
(1+n_{0}^{-1})-1}\,\end{equation} with high resolution in the
long-wavelength part of the spectrum, which is the most important and
informative in the context of BEC. The magnon energy was determined from
the frequency $\nu=\nu_{\mathbf{H}}(1+\omega)$, and the number of magnons
from the intensity of the emission in processes of inelastic Raman
scattering of a light wave on the magnon distribution established in the
film.

On the right-hand side of (\ref{3}) we have introduced the spectral
density of states
\begin{equation}\label{4} g
(\omega)=\sum_{\mathbf{k}}\delta(\omega-\omega_{\mathbf{k}}).\end{equation}
Because of the large difference in the longitudinal and transverse
dimensions of the film, the ratio of the corresponding parameters
$\omega_{j}$ is also large:
$\omega_{x}/\omega_{y}=\omega_{x}/\omega_{z}\sim10^{6}$. This leads to the
circumstance that the spectrum of states is split into layers around the
harmonics corresponding to the first component of the quasimomentum. This
peculiar structure of the spectrum allows one to decompose $g(\omega)$
and, hence, $n(\omega)$ into three characteristic terms with different
singular behavior in the low-frequency (near $\nu_{\mathbf{H}}$) region:
\begin{equation}\label{5} g
(\omega)=g_{C}(\omega)+g_{1}(\omega)+g_{\infty}(\omega)\,, \end{equation}
where
\begin{equation}\label{6}
g_{C}(\omega)=\delta(\omega-\omega_{0})\end{equation} is the term
corresponding to the contribution in (\ref{4}) from the lowest magnon
state; \begin{equation}\label{7} g_{1}(\omega)={\sum_{\
\mathbf{k}_{\perp}}}'\delta(\omega-\omega_{x}-\omega_{y}
\mathbf{k}_{\perp}^{2})\,, \quad
\mathbf{k}_{\perp}^{2}=k_{y}^{2}+k_{z}^{2}\,,\end{equation} is the
contribution of the first excited layer (the fundamental harmonic of the
first component of the quasimomentum) and, finally,
\begin{equation}\label{8}
g_{\infty}(\omega)=\sum_{k_{x}=2}^{\infty}\sum_{\ \,
\mathbf{k}_{\perp}}\delta
(\omega-\omega_{x}k_{x}^{2}-\omega_{y}\mathbf{k}_{\perp}^{2})\,\end{equation}
is the contribution of all the other states. The prime on the summation
sign in (\ref{7}) means that the term with $\mathbf{k}_{\perp}^{2}=2$ is
excluded, as it is taken into account explicitly in (\ref{6}). Changing
from sums to integrals in (\ref{7}) and (\ref{8}), we obtain
\begin{eqnarray}\label{9}
g_{1}(\omega)&\simeq&\frac{\pi}{2}\int\limits_{\sqrt{\varkappa+2}}^
{\infty}
dk\,k\,\delta(\omega-\omega_{x}-\omega_{y}k^{2})=\frac{\pi}{4\omega_{y}}
\theta(\omega-\omega_{x}-\varkappa\omega_{y})\,,\\\nonumber
g_{\infty}(\omega)&\simeq&\frac{\pi}{2}\sum_{k_{x}=2}^{\infty}
\int\limits_{0}^{\infty}dk\,k\,
\delta(\omega-\omega_{x}k_{x}^{2}-\omega_{y}k^{2})=\frac{\pi}
{4\omega_{y}}
\sum_{k_{x}=2}^{\infty}\theta(\omega-\omega_{x}k_{x}^{2})=\\&=&\frac{\pi}
{4\omega_{y}}
\biggl(\sqrt{\frac{\omega}{\omega_{x}}}-\biggl\{\sqrt{\frac{\omega}
{\omega_{x}}}
\biggr\}-1\biggr)\theta(\omega-4\omega_{x})\,,\label{10}\end{eqnarray}
where $0<\varkappa<3$, and the curly brackets on the right-hand side of
(\ref{10}) denote the fractional part of the quantity enclosed in them
(i.e., $\{s\}$ is the fractional part of the real variable $s$). Since
$0<\{s\}<1$, one can set the mean value of $\{s\}\simeq1/2$, which gives
for (\ref{10})
\begin{equation}\label{11} g_{\infty}(\omega)=\frac{\pi}{4\omega_{y}\sqrt{\omega_{x}}}
\left(\sqrt{\omega}-
\frac{3}{2}\sqrt{\omega_{x}}\right)\theta(\omega-4\omega_{x})\,.\end{equation}
The partial contribution of $g_{1}(\omega)$ (\ref{9}) is proportional to
the surface area of the sample, while in $g_{\infty}(\omega)$ (\ref{11})
the first term is proportional to the volume and the second to the area.
Collecting the corresponding terms, we write $g (\omega)$ in the following
form:  \begin{equation}\label{5a} g
(\omega)=g_{C}(\omega)+g_{S_{V}}(\omega)+g_{V}(\omega)\,, \end{equation}
where
\begin{eqnarray}\label{13a} g_{C}(\omega)&=&\delta(\omega-\omega_{0})\,,\\
\label{14a} g_{S_{V}}(\omega)&=&\frac{\pi}{4\omega_{y}}\ [\theta(\omega
-\omega_{0}-\varkappa\omega_{y})-\frac{3}{2}\,\theta(\omega
-4\omega_{x})]\,,\\\label{15a} g_{V}(\omega)&=&\frac{\pi
\sqrt{\omega}}{4\omega_{y}\sqrt{\omega_{x}}}\ \theta(\omega
-4\omega_{x})\,.\end{eqnarray} It is easily checked that for large
$\omega$ the phase volume element
$$d\Phi=[g_{V}(\omega)+g_{S_{V}}(\omega)]d\omega$$ agrees with the
asymptotic estimate (\ref{13d}. However, in the low-energy (threshold )
region of interest to us, the question of the lower limits of integration
over the phase volume becomes fundamental, and the answer to this question
is contained in expressions (\ref{13a})--(\ref{15a}).

In sum, by substituting the formulas obtained for $g(\omega)$ into
(\ref{3}), we get
\begin{equation}\label{12} n
(\omega)=n_{C}(\omega)+n_{S_{V}}(\omega)+n_{V}(\omega)\,,\end{equation}
where, in accordance with decomposition (\ref{5a}), the functions
\begin{eqnarray}\label{13}
 n_{C}(\omega)&=&n_{0}\delta(\omega-\omega_{0})\,,\\ \label{14}
n_{S_{V}}(\omega)&=&\frac{\pi}{4\omega_{y}}\ \frac{\theta(\omega
-\omega_{0}-\varkappa\omega_{y})-(3/2)\,\theta(\omega
-4\omega_{x})}{\bigl(1+n_{0}^{-1}\bigr) \exp(\frac{\omega-\omega_{0}}{t})
-1}\,,\\\label{15}
n_{V}(\omega)&=&\frac{\pi}{4\omega_{y}\sqrt{\omega_{x}}}\
\frac{\sqrt{\omega}\ \theta(\omega
-4\omega_{x})}{\bigl(1+n_{0}^{-1}\bigr)\exp(\frac{\omega-\omega_{0}} {t})
-1} \,,\end{eqnarray} specify the partial frequency distributions
corresponding to the condensate, surface, and volume contributions. In a
restricted range of frequencies $\omega$ not exceeding a certain value
$\omega_{max}$, expressions (\ref{14}) and (\ref{15}) simplify if
$n_{0}\gg1$ and $t\gg\omega_{max}$:
 \begin{eqnarray}\label{16} n_{S_{V}}(\omega)&=&\frac{\pi
t}{4\omega_{y}}\
\frac{\theta(\omega-\omega_{0}-\varkappa\omega_{y})-(3/2)\,\theta(\omega
-4\omega_{x})} {\omega-\omega_{0}+t/n_{0}} \,,\\\label{17}
n_{V}(\omega)&=&\frac{\pi t}{4\omega_{y}\sqrt{\omega_{x}}}\
\frac{\sqrt{\omega}\ \theta(\omega -4\omega_{x})}
{\omega-\omega_{0}+t/n_{0}}\,.\end{eqnarray} It is seen from these
expressions that because of the small parameter $\sqrt{\omega_{x}}$ in the
denominator of (\ref{17}), the contribution (\ref{16}) from the surface
term in the general case is negligible: \\ $n_{V}(\omega)\gg n_{S_{V}}
 (\omega)$. But at the
thresholds ($\omega=\omega_{0}+\varkappa\omega_{y}$ and
$\omega=4\omega_{x}$ for (\ref{16}) and (\ref{17}), respectively) the
situation is the opposite: $n_{S_{V}}\gg n_{V}$.

In real experiments the results of measurements depend on the resolving
power of the device, and usually one observes not $n (\omega)$ but the
spectral density averaged over some frequency interval $\delta\omega\simeq
Q^{-1}$:
\begin{equation}\label{18}
 n^{obs} (\omega)=
\int\limits_{-\infty}^{\infty}d\omega'n (\omega')
F(\omega-\omega')\,,\quad \int\limits_{-\infty}^{\infty}d\omega
F(\omega)=1,\end{equation} where $F(\omega)$  is the amplitude√frequency
characteristic (AFC) of the receiver. Because the value of $n_{C}(\omega)$
in (\ref{13}) is proportional to a $\delta$ function, the observable
condensate contribution
\begin{equation}\label{19} n^{obs}_{C}(\omega)=n_{0}F(\omega-\omega_{0})\end{equation}
actually reproduces the AFC. Expressions (\ref{18}) and (\ref{19}) show
that the formation of the resonance peak as the BEC signal involves the
participation not of just one state but also of other states whose
energies are close to the gap $\varepsilon_{\mathbf{H}}$. Therefore, at
some distance from the ``condensate'' frequency $\omega_{0}$ and not only
at the thresholds, the partial contributions $n^{obs}_{S_{V}}(\omega)$ and
  $n^{obs}_{V}(\omega)$ can have values comparable to
$n^{obs}_{C}(\omega)$, despite the quantitative difference in the
coefficients in expressions (\ref{16}) and (\ref{17}). Furthermore, owing
to the finite width of the AFC the observable magnon spectral density
$n^{obs}(\omega)$ is nonzero even in the region below the threshold
($\omega<0$, ХКХ $\nu<\nu_{\mathbf{H}}$). As the AFC we consider the
often-encountered function\footnote[5]{In radiophysics, for example, it
corresponds to a circuit consisting of a single oscillatory loop.}
\begin{equation}\label{20}
F(\omega)=\frac{1}{\pi}\,\frac{Q}{1+Q^{2}\omega^{2}}\,.\end{equation}
where $Q$  is the quality factor. The integrals (\ref{18}) of the
functions (\ref{16}), (\ref{17}), and the Lorentzian (\ref{20}) are taken
explicitly. Taking into account the smallness of the parameters
$\omega_{x}$, $\omega_{y}$ and $t/n_{0}$ and assuming that
$Q\omega_{0}\ll1$, we arrive at the expressions
\begin{eqnarray}\label{22} n^{obs}
(\omega)&=&n^{obs}_{C}(\omega)+n^{obs}_{V}(\omega)+
n^{obs}_{S_{V}}(\omega)\,,\\ \label{23}
n^{obs}_{C}(\omega)&=&\frac{n_{0}}{\pi}\,\frac{Q}{1+Q^{2}\omega^{2}}\,,\\
\label{24} n^{obs}_{V}(\omega)&=&\frac{\pi\,
t}{4\sqrt{2}\omega_{y}\sqrt{\omega_{x}}}\biggl(\frac{Q^{2}\omega}{1+Q^{2}
\omega^{2}}+ \frac{Q}{\sqrt{1+Q^{2}\omega^{2}}}\biggr)^{1/2}\,,\\
n^{obs}_{S_{V}}(\omega)&=&\frac{t}{8\omega_{y}}\,\frac{Q}{1+Q^{2}\omega^{2}}
\biggl[ A-Q\omega\biggl(\frac{\pi}{2}+\arctg Q\omega\biggr)
-\frac{1}{2}\ln(1+Q^{2}\omega^{2})\biggr]\,,\label{25} \end{eqnarray}
where $A$ denotes the $\omega$-independent quantity (which depends on $T$,
$n_{0}$, and the linear dimensions of the system)
\begin{equation}\label{26}
A=3\ln(t/n_{0}+3\omega_{x})-2\ln(t/n_{0}+\varkappa\omega_{y})+\ln
Q\,.\end{equation} It can be seen from expressions (\ref{23})--(\ref{26})
that the differential components of the spectral density behave
differently upon variation of the temperature, film thickness, and the
pumping done to produce magnons in sufficient number for BEC in the
system. This satisfies the basic prerequisites for reliable separation of
the condensate (coherent) component and the surface and volume
(incoherent) components of the observed total (and, in essence, unified)
spectral curve.

It is extremely significant that in its main details the shape of the
spectrum does not depend separately on the temperature and pumping but on
their ratio $t/n_{0}$. Expressions (\ref{23})--(\ref{26}) show that the
shape of the  $n^{obs}(\omega)$ curve is invariant with respect to
simultaneous variation of temperature and the number of particles in the
condensate, provided that their ratio $\eta=t/n_{0}$ is constant: at a
fixed value of $\eta$ variation of the temperature or magnon pumping
affects only a coefficient that is common to all contributions.

The critical magnon density $\mathfrak n_{\rm BEC}$ (the derivative of the
specific heat as a function of $\mathfrak n$) jumps at the transition
through $\mathfrak n_{\rm BEC}$)
 has the form
\begin{equation}\label{26'} \mathfrak n_{\rm
BEC}=\frac{\zeta(3/2)}{\lambda^{3}_{T}}\,.\end{equation} At room
temperature $T=300\,K$ and $m_{m}=5\,m_{e}$ the thermal length (see
(\ref{121d}) is small: $\lambda_{T}\simeq1.92\times10^{-7}$. Therefore the
number of particles in the condensate is already quite large:  (ОКНРМНЯРЭ
ЙНМДЕМЯЮРЮ $N_{\rm BEC}/V\simeq6.28\times10^{14}$, and the parameter
$\eta_{\rm BEC}=t/N_{\rm BEC}\simeq2.9\times10^{-8}$ is sufficiently small
as to ensure the validity of the approximations made in the derivation of
(\ref{23})--(\ref{26}). Despite the large number of particles in the
condensate, its contribution to $n^{obs}(\omega)$ is practically
unnoticeable (roughly three or four orders of magnitude smaller than the
thermal excitations). Nevertheless, please note that the $n^{obs}
 (\omega)$ curve
already shows a completely formed peak as a consequence of the singular
(at $\eta\,,\omega\rightarrow0$) behavior of the surface (\ref{16}) and
volume (\ref{17}) contributions to $n^{obs}(\omega)$. For direct
observation of $n^{obs}_{C}(\omega)$ in the spectral density it is
necessary that $n_{0}\gg N_{\rm BEC}$.

 \begin{figure}[h] \begin{center}
 \includegraphics[height=120mm,keepaspectratio=true]
 {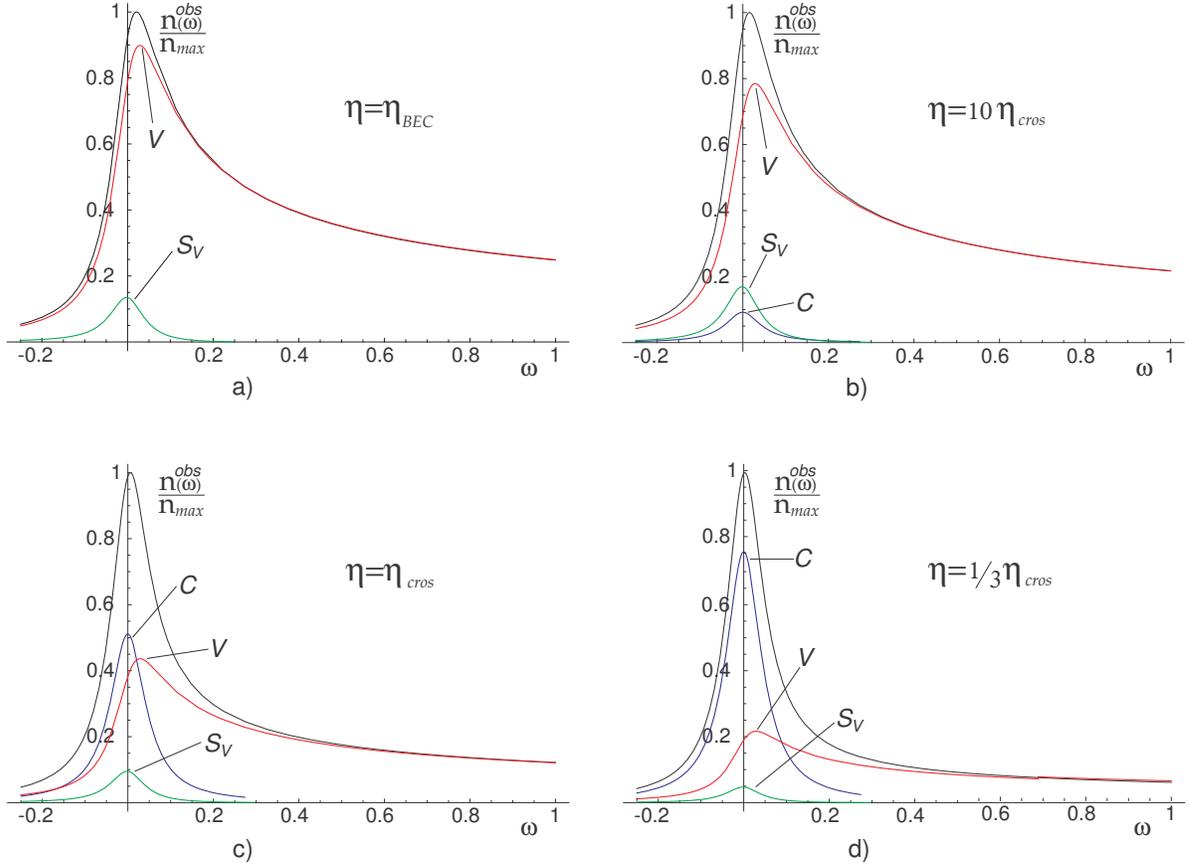}
 \caption{\small \ Spectral density of magnons and its partial contributions
 at different values of the
parameter $\eta$; $n_{max}$ is the normalized coefficient corresponding to
the maximum of the total spectral density at the given $\eta$;
$L_{x}=10~\mu m$, $\eta_{\rm BEC}\simeq4.8\times10^{-9}$, $\eta_{cros}
 \simeq1.92\times10^{-12}$; the symbols $C,\, S_{V}$,  and $V$ denote the corresponding partial
contributions.}
 \end{center} \end{figure}
\begin{figure}[h] \begin{center}
 \includegraphics[height=120mm,keepaspectratio=true]
 {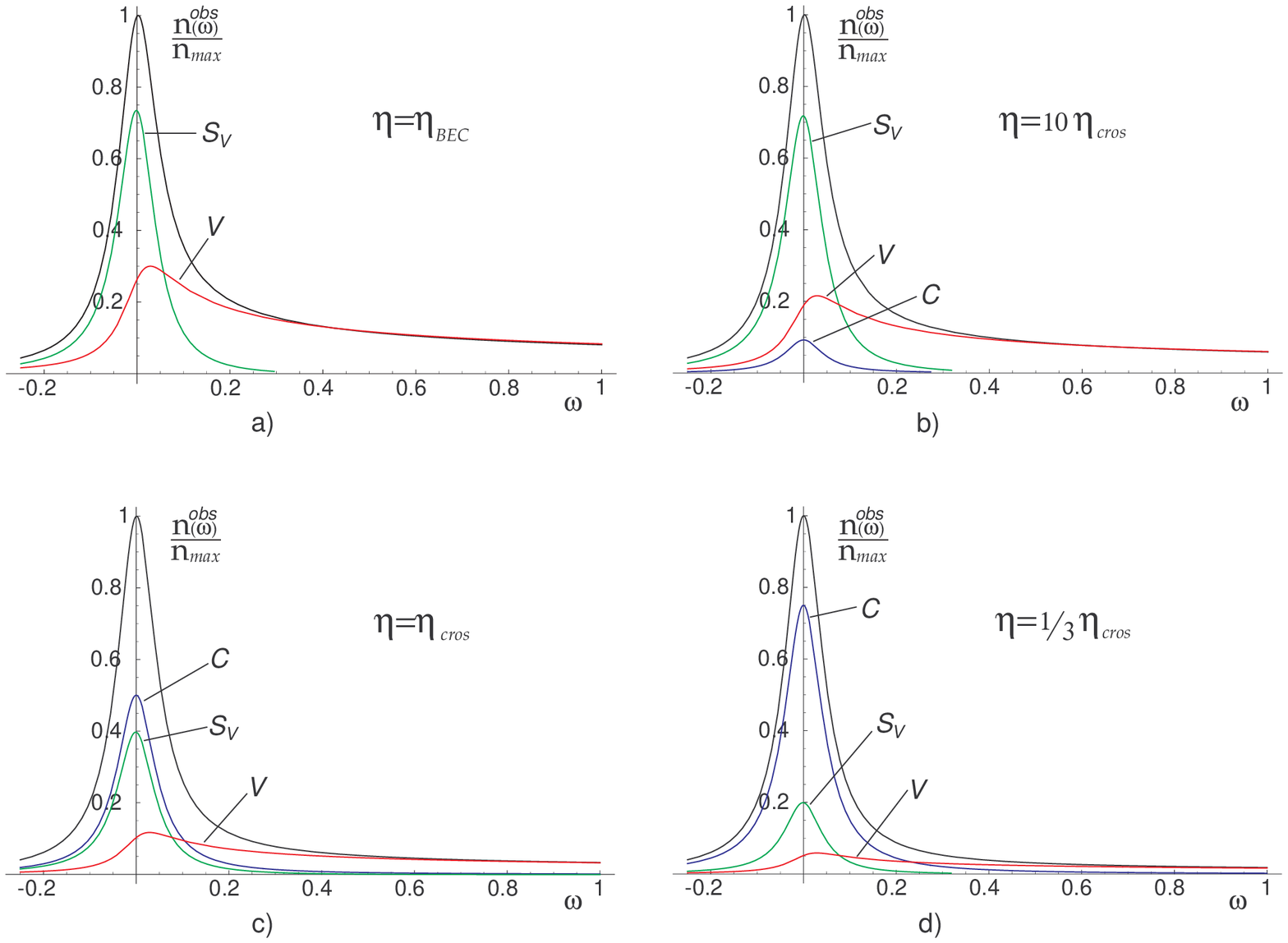}
 \caption{\small \ The same as in Fig. 3, for $L_{x}=1~\mu m$;
 $\eta_{\rm
BEC}\simeq2.22\times10^{-8}$, $\eta_{cros}\simeq5.26\times10^{-12}$.}
 \end{center} \end{figure}

It can be determined from (\ref{23})--(\ref{26}) that at such values of
the parameter $\eta=\eta_{cros}$ the number of particles in the condensate
becomes macroscopic --- e.g., equal to the volume contribution of the
thermalized excitations. This gives a rough but simple estimate:
  \begin{equation}\label{28}
\eta_{cros}=\frac{4\omega_{y}}{\pi^{2}}\sqrt{2Q\omega_{x}}\,.\end{equation}
For a film thickness $L_{x}=10\mu m$ this gives
$\eta_{cros}\simeq2.56\times10^{-12}$. When the parameter $\eta$ reaches
the value $\eta_{cros}$ (with increasing magnon pumping) a kind of
crossover occurs, i.e., the peak on the $n^{obs} (\omega)$ curve rises up
sharply. A more accurate estimate for $\eta_{cros}$ can be obtained from
the quantity $n^{obs}_{C}(0)=n^{obs}_{ex}(\omega_{max})$, where
$n^{obs}_{ex}(\omega)=n^{obs}_{V}(\omega)+n^{obs}_{S_{V}}(\omega)$ and
$\omega_{max}$ is the frequency at which the function
$n^{obs}_{ex}(\omega)$ reaches a maximum. Hence we obtain
$\eta_{cros}\simeq1.92\times10^{-12}$ for $L_{x}=10\mu m$, and
$\eta_{cros}\simeq5.26\times10^{-12}$ for $L_{x}=1\mu m$.

>From these estimates one is convinced that at room temperature,
$t=3\cdot10^{3}$, this crossover occurs at $n_{0}\simeq1.56\times10^{15}$
or a density $n_{0}/V\approx1.56\times10^{18}\,{cm}^{-3}$, a value which
has been attained in YIG films \cite{Demo}.

The above-discussed details of the behavior of the magnon spectral density
as a function of $n_{0}$ (which corresponds to the dependence on the
pumping level) are illustrated in Figs.~3--5. Figure 3 shows the spectrum
corresponding to a microfilm (investigated in \cite{Demo}) of thickness
$\approx10 \mu m$, and Fig.~4 to a film an order of magnitude thinner,
$\approx1 \mu m$. It is seen that in the thicker film at all pumping
levels the surface contribution is less than the volume contribution, and,
essentially, one can speak of only the competition of the condensate and
volume contributions. The latter of these, being asymmetric with respect
to the maximum and rather strong in the region of the short-wavelength
wing of the spectrum, weakens, yielding its place to the condensate
contribution that ultimately determines the shape of the band at the
highest pumping level (Fig. 3d).

More interesting, perhaps, is the picture of the formation of the spectrum
in a very thin film (Fig.~4), when all three contributions are present and
competing. Moreover, the surface contribution, as we see, even at
relatively weak pumping (Fig. 4a,b) not only remains larger but is
essentially dominant, and together with the rising condensate contribution
it largely determines the observed shape of the curve. Unlike the case
shown in Fig.~3, here the volume contribution, which is also asymmetric,
is relatively small (especially at large $n_{0}$), and the condensate
contribution mainly ``combats'' the surface contribution, prevailing only
at the highest pumping levels (Fig. 4d). Figures 3 and 4 clearly attest to
the possibilities opened up in the study of BEC not only in thin magnetic
films but also in other systems of finite size and with a small number of
particles.

Finally, Fig.~5 shows the evolution of the total distribution as the
temperature is varied, leading, under otherwise equal conditions, to
substantial growth of the number of magnons in the condensate with
decreasing temperature.
\begin{figure}[h] \begin{center}
 \includegraphics[height=60mm,keepaspectratio=true]
 {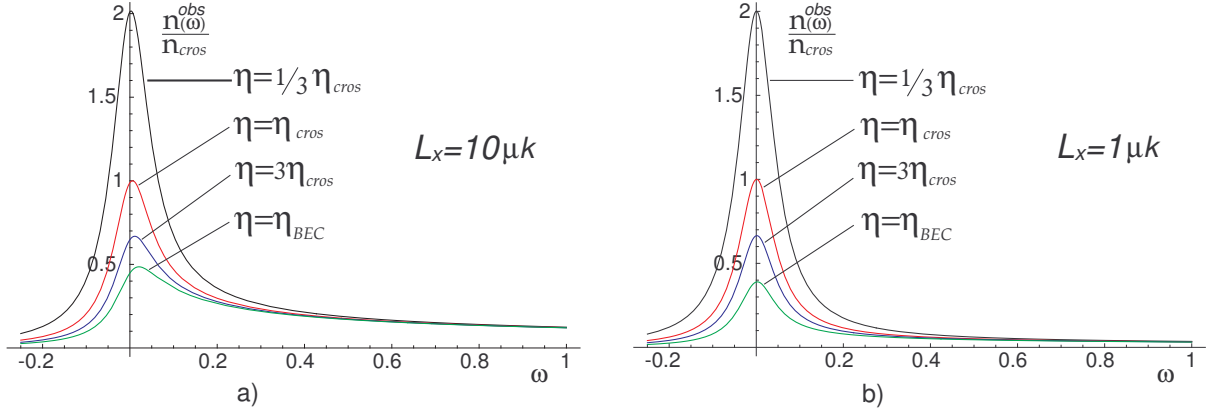}
 \caption{\small \ Total spectral density of magnons at different values of the parameter
 $\eta$. $n_{cros}$ is the
normalized coefficient corresponding to the maximum of the total spectral
density at  $\eta=\eta_{cros}$.}
 \end{center} \end{figure}
 It should be kept in mind that the crossover point
$\eta_{cros}$ here cannot be interpreted as a phase transition signal. In
the dimensionless variables used, the temperature of the latter,
(\ref{1v}), is given by the expression
  \begin{equation}\label{29} t_{\rm
BEC}=\frac{k_{B} T_{\rm
BEC}}{\varepsilon_{\mathbf{H}}}=\frac{4}{\pi}(\omega_{x}\omega_{y}
\omega_{z})^{1/3}
\biggl[\frac{N}{\zeta(3/2)}\biggr]^{2/3}\,.\end{equation} As was shown
above (see Eq. (\ref{26'})), the number of particles of the condensate at
the phase transition point is $N_{\rm BEC}\sim N ^{2/3}$, and so
\begin{eqnarray}\nonumber \eta_{\rm BEC}&=&\frac{t_{\rm BEC}}{n_{\rm
BEC}}=\frac{4}{\pi}\,\frac{(\omega_{x}\omega_{y}
\omega_{z})^{1/3}}{\zeta^{2/3}(3/2)}\,,\\
\frac{\eta_{cros}}{\eta_{\rm BEC}}&=&\frac{\zeta^{2/3}(3/2)}{\pi}\sqrt{2Q}
(\omega_{x}\omega_{y}\omega_{z})^{1/6}\ll1.\label{30}\end{eqnarray} This,
in turn, indicates that the appearance of a noticeable peak in the
spectral density essentially occurs already in the BEC phase. For the
present case of a film with $L_{x}\simeq10\mu m$, $L_{y}=L_{z}\simeq1$\,cm
and for $Q\simeq20$ the ratio $\eta_{cros}/\eta_{\rm BEC}=N_{\rm
BEC}/n_{cros}\simeq4\cdot10^{-4}$. In other words, at a fixed temperature
a significantly lower density is necessary for a phase transition to the
state with the condensate than for observation of the crossover, i.e., the
transition to the state with a predominant number of particles in the
condensate (see Figs.~3 and 4).

It should be emphasized that the contribution $n^{obs}_{S_{V}}(\omega)$
(\ref{25}) to the spectral density is a mesoscopic or finite-size effect
that is isolated in analytical form. One notices that the frequency
dependence of this contribution is practically no different from
$n^{obs}_{C}(\omega)$. Meanwhile, the physical nature of these
contributions is different: $n^{obs}_{C}(\omega)$ is just the contribution
of a large number of magnons with identical quantum numbers. This set of
quasiparticles is a {\it quantum object} consisting of a macroscopic
number of particles, or a coherent Bose condensate. On the contrary,
$n^{obs}_{S_{V}}(\omega)$ is formed of a set of magnons with different
quantum numbers and corresponds to an obviously incoherent state.

It is scarcely possible to distinguish these contributions by the
attribute of coherence with the aid of the corresponding interference
measurements. Nevertheless, because the contribution
$n^{obs}_{S_{V}}(\omega)$ is proportional to the temperature and the
surface area of the sample [unlike $n^{obs}_{C}(\omega)$], these
components of the spectral density can be distinguished with the aid of a
series of measurements at different temperatures on samples of different
size and shape.

For a quantitative description of the data it is necessary, among other
things, to have information on the effective AFC of the measuring setup.
The Lorentzian (\ref{20}) that we used as a model can scarcely correspond
to reality. Nevertheless, even in this simple model one sees a substantial
dependence of the results of observation on the parameters of the AFC, in
the present case---on the quality factor $Q$. Considering the partial
contributions as functions of the reduced frequency $Q\,\omega$, it is
easy to see that the contribution of the condensate goes as $\sim Q$,
while the volume contribution (\ref{24}) goes as $\sim Q^{1/2}$.
Therefore, with increasing resolving power the contribution of the
condensate becomes more and more noticeable against the background of the
volume component. This can be used for identification of the volume
component. However, the surface contribution is also practically
proportional to the  factor, and therefore variation of the $Q$ factor is
insufficient for separating it from the condensate component, and special
data processing is necessary (such as that mentioned above).

\section{Inclusion of anisotropy of the magnon dispersion relation}

As we have said, in real ferromagnetic films, because of the contribution
of the magnetic dipole interaction, the spectrum of long-wavelength
magnons is anisotropic. Here the anisotropy is determined by the external
magnetic field, which sets the magnetization direction \cite{Gur,Demo}.
Taking that interaction into account leads to a distortion of the
isotropic quadratic dispersion relation (\ref{1}) such that its dependence
on the third component of the quasimomentum exhibits a characteristic dip
(see Fig.~6) at a value $p_{z}=p_{0}\simeq3\cdot10^{-4}\,cm^{-1}$ (recall
that ${\mathbf{H}}||z$).
\begin{figure}[h]
 \begin{center}
 \includegraphics[height=60mm,keepaspectratio=true]
 {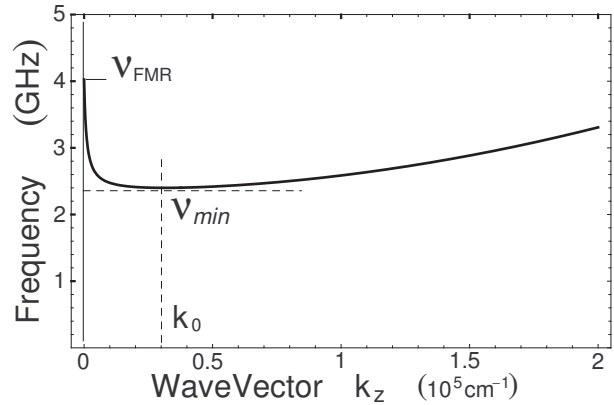}
 \caption{\small  \  Form of the dispersion relation of long-wavelength magnons with wave vectors
 $\mathbf{k}||\mathbf{H}$ in a ferromagnetic microfilm.}
 \end{center} \end{figure} The corresponding $\varepsilon({\mathbf{p}})$ dependence is well
approximated by the  expression
\begin{equation}\label{1t}
\varepsilon(\mathbf{p})=\frac{{\mathbf{p}}^{2}}{2m}+\frac{\varepsilon(0)}
{p_{z}/p_{a}+1}\,,\end{equation} which for $p_{a}\rightarrow\infty$ and
$\varepsilon (0)\rightarrow\varepsilon_{\mathbf{H}}$ goes over to
(\ref{1a2}). If one knows the energy at zero, $\varepsilon(0)$, and the
parameters of the extremum, $p_{0}$ and
$\varepsilon_{0}\equiv\varepsilon(p_{0})$, then the momentum $p_{a}$ on
the right-hand side of (\ref{1t}) can be written in the form
 \begin{equation}\label{2t} \frac{p_{0}}{p_{a}}=\frac{\varepsilon(0)}
{\varepsilon_{0}-p_{0}^{2}/2m }-1.\end{equation} In accordance with the
phenomenological curve in Fig.~6 we have:
$\varepsilon(0)/2\pi\hbar\simeq4$GHz,
$\varepsilon_{0}/2\pi\hbar\simeq2.3$GHz. However, for the sake of
unification in the case considered above we assume that the energy at the
minimum, $\varepsilon_{0}$, corresponds to the gap energy
$\varepsilon_{\mathbf{H}}$ of the previous Section
($\varepsilon_{0}/2\pi\hbar\simeq2$GHz). Then Eq. (\ref{1t}) is written in
dimensionless variables as
 \begin{eqnarray}
\varepsilon({\mathbf{p}})&\equiv&\varepsilon_{\mathbf{k}}=
\varepsilon_{0}(1+\omega_{\mathbf{k}}),\quad \omega_{\mathbf{k}}=
\omega_{x}k_{x}^{2}+\omega_{y}k_{y}^{2}+\omega_{z}\,f(k_{z}),\nonumber\\
f(k_{z})&=& (k_{z}-k_{0})^{2}\biggl(1+\frac{2k_{0}}{k_{z}+k_{a}}
\biggr)=k_{z}^{2} +\frac{2k_{0}(k_{0}+k_{a})^{2}}{k_{z}+k_{a}}
-k_{0}(3k_{0}+2k_{a}), \label{3t}\end{eqnarray} where
$$k_{0}\simeq10^{4},\quad k_{a}\simeq270.$$ We note that the minimum of energy in (\ref{3t})
 corresponds to the state with quantum numbers
k$\mathbf{k}=(1,1,k_{0})$.

Calculation of the spectral density of the number of states
$$g(\omega)=\sum_{{\mathbf{k}}}\delta(\omega-\omega_{\mathbf{k}})$$ with
the anisotropic dispersion relation (\ref{3t}) is complicated considerably
in comparison with the simple calculations in Sec. III. Nevertheless, the
final expressions are quantitatively very close to those obtained for the
isotropic case. Repeating the calculation scheme of the previous Section,
we separate out from $g(\omega)$ the lowest state $g_{C}(\omega)$, the
contribution of the first ($(k_{x}=1)$) harmonic $g_{1}(\omega)$, and the
remaining part $g_{\infty}(\omega)$:
$$g(\omega)=g_{C}(\omega)+g_{1}(\omega)+g_{\infty}(\omega),$$
$$g_{C}(\omega)=\delta(\omega-\omega_{x}-\omega_{y}),$$
\begin{eqnarray}\nonumber g_{1} (\omega)&=&
{\sum_{{\mathbf{k}}_{\perp}}}'\delta[\omega-\omega_{x}-\omega_{y}k_{y}^{2}
-\omega_{z}f(k_{z})]\simeq\\\label{4t}&\simeq&
\frac{J(\omega)}{2\sqrt{\omega_{y}\omega_{z}}}\,\theta(\omega-\omega_{x}-
\omega_{y})\,. \end{eqnarray} The notation $J(\omega)$ on the right-hand
side of (\ref{4t}) stands for the integral (which is evaluated in Appendix
B)
\begin{equation} J(\omega)=\int
\limits_{z_{1}}^{z_{2}}\frac{dz}{\sqrt{\omega-\omega_{x}-w(z)}}\,,
\label{5t0}\end{equation} where $z_{1}$ and $z_{2}$ are roots of the
equation  $w(z)=\omega-\omega_{x}$:
\begin{equation}\label{5t}
w(z)=\omega_{z}f(z/\sqrt{\omega_{z}})=(z-z_{0})^{2}
\biggl(1+\frac{2z_{0}}{z+z_{a}}\biggr),\end{equation} where $z_{a}=
k_{a}\sqrt{\omega_{z}}\simeq2.7\cdot10^{-3}$,
$z_{0}=k_{0}\sqrt{\omega_{z}}\simeq10^{-1}$.

In an analogous way, by calculating the contribution $g_{\infty}(\omega)$
to accuracy $O(\sqrt{\omega_{x}})$, we find
  \begin{eqnarray}\nonumber g_{\infty}(\omega)&=&
 \sum_{k_{x}=2}^{\infty}
 \sum_{\ \ {\mathbf{k}}_{\perp}}\delta[\omega-\omega_{x}k_{x}^{2}-
 \omega_{y}k_{y}^{2}-\omega_{z}f(k_{z})]\simeq\\\label{8t}&\simeq&
 \frac{\theta(\omega-4\omega_{x}-\omega_{y})}{2\sqrt{\omega_{x}\omega_{y}
 \omega_{z}}}\,\biggl[\frac{\pi}{2}(z_{2}-z_{1})
 -\frac{3}{2}\sqrt{\omega_{x}}J(\omega)\biggr]\,.\end{eqnarray}
 Collecting terms proportional to the surface and volume in (\ref{4t}) and (\ref{4t}), we obtain
(cf. (\ref{14a}) and (\ref{15a})): \begin{eqnarray} \label{10t}
g_{S_{V}}(\omega)&=&\frac{J(\omega)}{2\sqrt{\omega_{y}\omega_{z}}}\
[\theta(\omega -\omega_{x}-\omega_{y})-\frac{3}{2}\,\theta(\omega
-4\omega_{x}-\omega_{y})]\,,\\\label{11t} g_{V}(\omega)&=&\frac{\pi~
(z_{2}-z_{1})}{4\sqrt{\omega_{x}\omega_{y}\omega_{z}}}\ \theta(\omega
-4\omega_{x}-\omega_{y})\,.\end{eqnarray}

It is easily checked that the integral $J(\omega)$, generally speaking,
depends weakly on $\omega$; for example, for $\omega\gg\omega_{x}$ the
function $J(\omega)\simeq\pi/2$, which coincides with our previous
investigation of the isotropic case. A quantitative difference, though
slight, appears only in the region of very low frequencies (near the
threshold), as is illustrated in Fig.~7.
\begin{figure}[h]
\begin{center}
\includegraphics[height=55mm,keepaspectratio=true]
{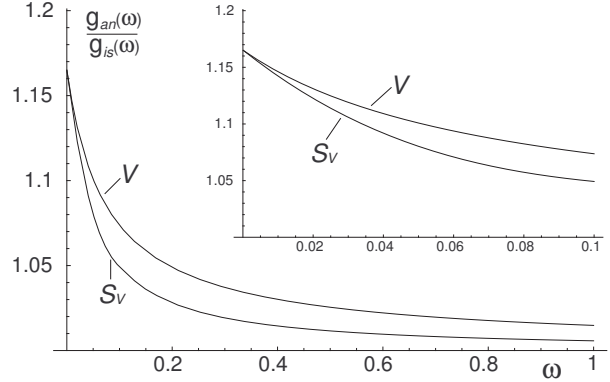} \caption{\small \ Ratio of spectral densities of states:
$g_{an}(\omega)$ corresponds to the anisotropic dispersion relation
(\ref{1t}), and $g_{is}(\omega)$ to the isotropic one (\ref{1a2}); as
above, the symbol $V$  denotes the volume contributions, $S_{V}$ the
surface contributions.}
\end{center} \end{figure}
\begin{figure}[h]
\begin{center}
\includegraphics[height=70mm,keepaspectratio=true]
{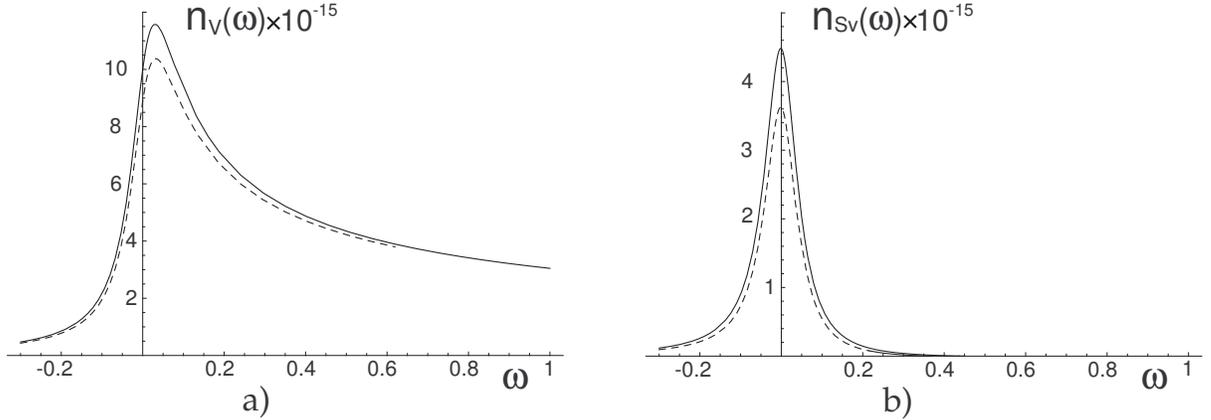} \caption{\small \ The volume (a) and surface (b) contributions
to the spectral density of magnons for a film 10~$\mu m$ thick; the solid
curve corresponds to the anisotropic dispersion relation (\ref{1t}), and
the dashed curve to the isotropic one (\ref{1a2}).}
\end{center} \end{figure}

Therefore, both the surface and volume contributions to the spectral
density of thermal spin excitations (\ref{8t}) are quantitatively similar
from the case of an isotropic dispersion relation (\ref{1a2}). On the
whole, the observed spectral distribution in this case is qualitatively
the same as the curves shown in Figs.~3 and 4.

The calculations done lead to a completely optimistic conclusion for
experiment: it would seem that the rather substantial difference in the
dispersion relations (\ref{1a2}) and (\ref{1t}) has almost no effect on
the results of measurement of the spectral density of magnons, as is
illustrated in Fig.~8. Thus the model of an ideal degenerate Bose gas with
isotropic quadratic spectrum is completely adequate for interpretation of
experiments on the BEC of magnons in real ferromagnetic systems.

\section{Conclusion}
The calculations done here convincingly demonstrate that in a thin
ferromagnetic film with dimensions of $1\,cm\times1\,cm\times10\,\mu m$
the BEC of spin excitations can be achieved already at a level of pumping
that ensures the appearance and existence of $\sim10^{15}$ magnons in the
film. The fundamental thing is that there are practically no restrictions
on temperature, and all the signs of BEC appear at rather high
temperatures, including room temperature. Of course, lowering the
temperature to the liquid hydrogen or helium region should lower the level
of the ``critical'' pumping in view of the fact that, as we have shown,
there is scaling or, in other words, the magnon spectral density depends
not on the total number of particles and temperature but only on their
ratio $N/T$. The shape of the observable spectrum of long-wavelength
magnons remains unchanged when this ratio is held constant. We recall that
the number of these quasiparticles is completely specified by the external
pumping.

On the other hand, the shape of the measured spectrum is largely connected
to the AFC of the filter and can also be varied by improving its $Q$. Its
increase (or the segregation of a narrower frequency region around the
lowest condensed magnon states) also leads to narrowing of the bandwidth,
and consequently, to a more definite investigation specifically of the
Bose condensate on whose consituent quasiparticles the Brillouin
scattering of the light wave occurs.

In interpreting the experimental data it should be recalled that one
cannot claim direct observation of the condensate solely on the basis that
a peak appears in the spectral density. All of the contributions have a
similar structure, and it is necessary to separate each of them reliably
from the experimental data. It follows from the theory that the main
distinguishing features of the condensate contribution is its independence
of temperature and the size of the sample. Therefore a precise and
convincing separation of $n^{obs}_{C}(\omega)$ requires additional
measurements at different temperatures.

In studying the BEC phenomenon in magnetic films, we have treated them as
systems of finite size. In such systems, as we have said, it is known that
BEC occurs not as a phase transition but as crossover. However, the
condensation, which occurs near a certain temperature $T_{cros}$, occurs
so rapidly that that temperature, though not a critical temperature, can
be treated as such. Apparently, the transition could be made more smooth
(or smeared in temperature) if an even thinner film were used while the
rather high measurement temperature was maintained. Here the absence of a
phase transition point in no way precludes the phenomenon of BEC, or the
macroscopic accumulation of magnons in their lowest energy state.
Moreover, thin films are interesting in still another respect: in them one
could trace the role and contribution of the correction terms (see Eq.
(2.14)) arising when one goes from summation to integration over phase
volume elements, something that, as far as we know, has hitherto escaped
attention.

The above calculations of the spectral density of magnons in a
ferromagnetic film presuppose the presence of an equilibrium magnon gas,
which, strictly speaking, does not correspond to reality. Intense
electromagnetic pumping creates relatively short-lived spin excitations,
as a result of four-magnon interaction, which, as was shown in \cite{Kal},
rapidly relax to a quasiequilibrium distribution with a temperature equal
or nearly equal to the temperature of the crystal (owing to the
spin√lattice coupling). Consideration of magnon√magnon and magnon√phonon
relaxation processes simultaneously with the process of magnon Bose
condensation undoubtedly requires a special analysis and will be done
separately.

Finally, we note that it has been proposed in an experimental paper
\cite{Demo} that the accumulation of magnons at two (symmetric) points of
the spin-wave spectrum fulfills the prerequisites for a nonuniform Bose
condensate. However, from the standpoint of the thermodynamics of the
process, no finite number of degenerate points of $\mathbf k$ space will
affect the observed scattering pattern and the spectral density
corresponding to it, which will be completely described in the framework
of the approach stated above. This conclusion remains in force, despite
the fact that the condensate that arises in such a case can be defined as
incoherent. A distinction would arise only in the case when the lowest
state of the magnon spectrum for some reason was infinitely degenerate
(e.g., this could correspond to a model spectrum in the form of a trough).
The condensate corresponding to that case is also incoherent, and the
transition to it would have some distinctions. However, the study of the
features of such a model is beyond the scope of this paper.

We are sincerely grateful to G.A.~Melkov for acquainting us with the
results of the experimental study by him and his coauthors, which
stimulated our study, and for numerous discussions and constructive
criticism.

This study was supported in part by the grant SCOPES N IB7320-110840 SMSF,
grant 02.07/00152 of the DFFD of Ukraine, and by the target program of the
Department of Physics and Astronomy of the National Academy of Sciences of
Ukraine.

\bigskip
  {\ \hspace{10cm}\underline{\bf APPENDIX A}}

\vspace{.5cm}

To demonstrate the problem with not separating the singular term, we
consider a series whose sum is expressed in terms of elementary functions:
$$S=\sum_{l=0}^{\infty}\frac{1}{l^{2}+x}=\frac{1}{2x}+\frac{\pi}{2}\,
\frac{\coth(\pi\sqrt{x})}{\sqrt{x}}\,. \eqno{({\text A}.1)}
$$ A ``naive'' transition from the sum to an integral in Eq. (A.1) gives the following approximation
for $S$:
$$S\approx S_{0}=\int\limits_{0}^{\infty}\frac{dl}{l^{2}+x}=\frac{\pi}
{2\sqrt{x}}\,. \eqno{({\text A}.2)}
$$ If the first term in Eq. (A.1), corresponding to $l=0$, is separated off, and the remaining
sum is approximated by an integral, we then get
$$S\approx S_{1}=\frac{1}{x}+\int\limits_{1/2}^{\infty}\frac{dl}{l^{2}+x}=\frac{1}{x}+
\frac{\arctan(2\sqrt{x})}{\sqrt{x}}\,, \eqno{({\text A}.3)}
$$ It is easy to check that approximation (A.3) is practically no different from the exact expression
(A.1) in the whole range of variation of the parameter $x>0$.
Approximation (A.2), on the contrary, is a very crude approximation
compared to (A.3), especially at $x\to0$. For example, one can compare
their asymptotic behavior at large and small $x$:
\begin{eqnarray*}
x\to0:\quad
S&=&\frac{1}{x}+\frac{\pi^{2}}{6}-\frac{\pi^{4}x}{90}+O(x^{2})\,,\\
S_{1}&=&\frac{1}{x}+2-\frac{8x}{3}+O(x^{2})\,,\\ x\to\infty:\quad
S&=&\frac{\pi}{2\sqrt{x}}+\frac{1}{2x}+O(\mbox{e}^{-2x})\,,\\
S_{1}&=&\frac{\pi}{2\sqrt{x}}+\frac{1}{2x}+\frac{1}{24x^{2}}+O(x^{3})\,.
\end{eqnarray*} Figure 9 shows graphs of $S$ as a function of the parameter $x$
for the exact and approximate expressions.
\begin{figure}[h] \begin{center}
 \includegraphics[height=55mm,keepaspectratio=true]
 {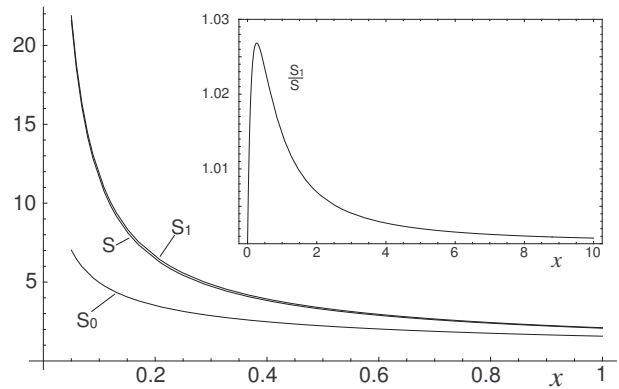}
 \caption{\small \ Behavior of the series $S$ and its approximations $S_{0}$ and $S_{1}$
 as functions of the parameter
$x$; the inset shows a plot of the ratio $S/S_{1}$.}
 \end{center} \end{figure}

\newpage
 {\ \hspace{10cm}\underline{\bf APPENDIX B}}

\vspace{.5cm}

Consider the integral (see (\ref{5t0}))
$$
J(\omega)=\int\limits_{z_{1}}^{z_{2}}\frac{dz}{\sqrt{\omega-w(z)}}\,,
\eqno{({\text B}.1)}$$ where
$$w(z)=(z-z_{0})^{2}\left(1+\frac{2z_{0}}{z+z_{a}}\right)\,.$$
The limits of integration in (B.1) are the positive roots (see Fig.~10) of
the equation $w(z_{1,2})=\omega$.
\begin{figure}[h]
 \begin{center}
 \includegraphics[height=60mm,keepaspectratio=true]
 {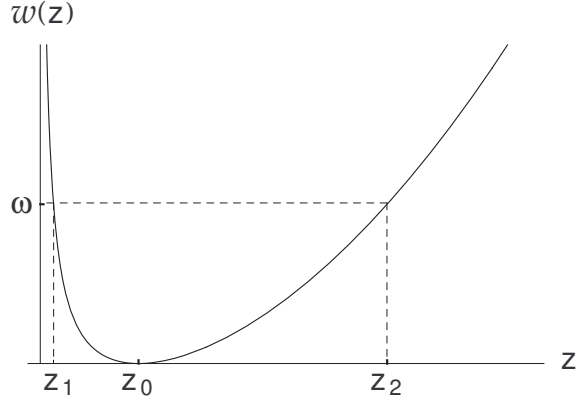}
 \caption{\small \ Graphical solution of the equation $w(z_{1,2})=\omega$.}
 \end{center} \end{figure}

For calculating the integral, the expression under the square root in
(B.1) is conveniently written in the form
 $$\omega-w(z)=\frac{(z_{2}-z)(z-z_{1})(z+z_{a}+z_{1}+z_{2})}{z+z_{a}}\,.$$
 Then with the standard change of the integration variable
 $$z=\frac{z_{1}(z_{2}+z_{a})+
 z_{a}(z_{2}-z_{1})\sin^{2}\varphi}{(z_{2}+z_{a})-(z_{2}-z_{1})
 \sin^{2}\varphi}$$
  integral (B.1) is transformed to $$
 J(\omega)=
 \frac{2(z_{a}+z_{1})}{\sqrt{(z_{a}+z_{2})(z_{a}+2z_{1}+z_{2})}}
 \int\limits_{0}^{\pi/2}\frac{d\varphi}{(1-a\sin^{2}\varphi)
 \sqrt{1-ab\sin^{2}\varphi}}\,,\eqno({\text
B}.2)$$ where
 $$a=\frac{z_{2}-z_{1}}{z_{a}+z_{2}}\,,\quad
 b=\frac{z_{2}+z_{1}}{z_{a}+2z_{1}+z_{2}}\,.$$ The integral on the right-hand side of (B.2)
 is the Legendre complete elliptic integral of the
third kind, $\Pi(a,ab)$.

\bigskip
\centerline{\bf NOTE ADDED IN PROOF}

\medskip
 After this article was submitted to
Low Temperature Physics, an experimental paper appeared [J. Kasprzak, M.
Richard, S. Kundermann, A. Baas, J. M. J. Keeling, F. M. Marchetti, M. H.
Szymanska, R. Andre, J. L. Staehli, V. Savona, P. B. Littlewood, B.
Deveaud, and Le Si Dang, Nature {\bf 443}, 409 (2006)] in which the
Bose√Einstein condensation of quasiparticles was also reported. In that
paper a different type of relatively light quasiparticles of the Bose type
was considered, specifically, exciton polaritons, a high density of which
can be produced by laser pumping in optical microcavities in a CdTe
crystal. In it, as in YIG, thermalization of the pumped quasiparticles
occurs, and above a certain quasiparticle density there is macroscopic
occupation of the quasiparticle (in this case, polariton) ground state, at
a rather high critical temperature $T_{\rm BEC}\approx19$~K. Evidence of
Bose condensation is provided by weak coherent correlation effects
observed in the radiation emitted from the crystal. There, however, it is
not ruled out that additional information about the formation of a Bose
condensate specifically could be obtained if the volume and surface
contributions to the observed intensity and line shape could be separated
(on the basis of their temperature or concentration behavior). Although in
our paper these contributions are calculated for magnons, they should
undoubtedly also be present for polaritons in the small volumes of the
optical microcavities.

\end{document}